\documentstyle[11pt]{article}
\title{The Electroweak Chiral Lagrangian and
  CP-Violating Effects in Technicolor Theories}
\author{Thomas Appelquist and Guo-Hong Wu\\
Department of Physics, Yale University, New Haven, CT 06520}
\date{June 28, 1994}
\begin{document}
\setlength{\baselineskip}{24pt}
\maketitle
\begin{picture}(0,0)(0,0)
\put(300,250){YCTP-P9-94}
\end{picture}
\vspace{-48pt}

\begin{abstract}

  We estimate the CP-violating $WW\gamma$ and $WWZ$ anomalous form
factors, arising from CP-violating interactions in extended
technicolor theories, and discuss their future experimental
detectability.  The electric dipole moment of the $W$ boson is found
to be as large as ${\cal O}(10^{-21}) \; \mbox{$e$ cm}$.  We connect
the CP-odd $WW\gamma$ and $WWZ$ couplings to the corresponding
CP-violating electroweak chiral lagrangian operators.  The electric
dipole moments of the neutron and the electron in technicolor theories
are estimated to be as large as ${\cal O}(10^{-26}) \; \mbox{$e$ cm}$
and ${\cal O}(10^{-29}) \; \mbox{$e$ cm}$ respectively.  We also
suggest the potential to observe large CP-violating technicolor
effects in the decay $t \rightarrow b + W^+$.
\end{abstract}

\section{Introduction}

   Recent work on extended technicolor (ETC) model building
\cite{ENKMRS,AT} has demonstrated that technicolor (TC) theories
could survive the challenges posed by precision electroweak
measurements, the fermion mass spectrum, and limits on
flavor-changing neutral currents and various rare decay modes.
However, CP violation in technicolor theories remains a puzzling
problem despite much study beginning more than ten years ago
\cite{ELPBDS}. In this paper we reexamine this problem in a generic
ETC model, with attention focused on the CP-odd $WW\gamma$ and $WWZ$
anomalous interactions and their induced effects.

   The electroweak symmetry breaking sector in technicolor theories
can be most conveniently parameterized below the breakdown scale of
chiral perturbation theory $\Lambda_{\chi}$ by a gauged nonlinear
chiral lagrangian \cite{ABL,EWR}. The scale $\Lambda_{\chi}$ is
approximately given by \cite{SS},
\begin{eqnarray}
 \Lambda_{\chi} \simeq \frac{4 \pi f}{\sqrt{N_f}} ,
\end{eqnarray}
 where $f = \frac{(\sqrt{2}G_F)^{- \frac{1}{2}}}{\sqrt{N_f/2}} \simeq
\frac{250}{\sqrt{N_f/2}}$ GeV is the Goldstone boson decay constant,
and where $N_f$ is the number of technifermion flavors ($N_f/2$ is
the number of technifermion weak doublets).  Restricting attention to
the minimal $SU(2)_L \times SU(2)_R$ chiral symmetry of the
technifermions and imposing electroweak gauge invariance, thirteen
CP-conserving and three CP-violating operators up to dimension-four
\footnote{Dimension counting of the electroweak chiral
lagrangian operators will be explained in section~2.}  can be written
down after using the classical equations of motion \cite{ABL,AW}.
The electroweak $S$, $T$ and $U$ parameters \cite{PT}, and the
CP-even and CP-odd $WW\gamma$ and $WWZ$ couplings \cite{Hag}, can all
be conveniently described by these operators of the electroweak
chiral lagrangian \cite{AW}.

   When extended to a larger global symmetry (as in the one family TC
model to be used in this paper), there will be pseudo-Goldstone-bosons
(PGBs) appearing as effective degrees of freedom of the electroweak
chiral lagrangian, and more operators can be written down beyond those
for the minimal $SU(2)_L \times SU(2)_R$ chiral symmetry.  These
additional operators will not, however, generate any new interactions
at tree level among the electroweak gauge bosons, beyond those already
existing in the gauged $SU(2)_L \times SU(2)_R$ chiral lagrangian.  In
this paper, we will concentrate on the anomalous interactions among
the electroweak gauge bosons. In particular, we will study the effects
of CP-violating, ETC interactions on the triple-gauge-boson vertices
(TGV's) measured below $\Lambda_{\chi}$. As the strength of the
vertices involving PGBs is expected to be the same as those involving
only electroweak gauge bosons, the contribution of the PGBs to the
CP-odd TGVs will be suppressed by at least $\frac{\alpha}{\pi}$
($\alpha$ is the fine structure constant) relative to the direct
contribution from the CP-violating ETC interactions. We will therefore
ignore the PGBs in this paper, and will restrict attention to the
gauged $SU(2)_L \times SU(2)_R$ electroweak chiral lagrangian.

  The size of the CP-violating dimension-four operators of the chiral
lagrangian depends on the details of the high energy ETC sector, and
simple dimensional estimates may not be trustworthy.  Besides the
suppression factor $\frac{\Lambda_{\chi}^2}{m_{ETC}^2}$ coming
naturally from the ETC sector ($m_{ETC}$ denotes the relevant ETC
scale), a specific model may introduce a small dimensionless parameter
into the final results if, for example, CP conservation is only
slightly violated in extended technicolor interactions.  At energies
above $\Lambda_{\chi}$ but below $m_{ETC}$, CP violation can be
described effectively by a set of ETC-induced, CP-violating
four-fermion operators.  These operators are added to the usual
technicolor lagrangian along with corresponding CP-conserving
four-fermion operators.  The technicolor interaction itself respects
CP symmetry.

  In this paper we estimate the effect of these CP-violating
four-fermion operators on the CP-odd TGVs, and on further amplitudes
induced by the CP-odd TGVs.  In the minimal standard model, the CP-odd
TGVs are expected to arise via the Kobayashi-Maskawa (KM) mechanism
\cite{KM} at the three-loop level \cite{EDMw}, and thus are suppressed
by at least two powers of the weak coupling $\frac{\alpha}{\pi}$ and
by the small mixing effect among the three families of quarks. They
are thus much too tiny to have experimental signatures in the near
future.  In technicolor theories, however, assuming maximal CP
violation in the ETC sector, our estimates show that these
CP-violating TGVs are much larger than in the minimal standard model,
and are comparable to or larger than in other extensions of the
standard model \cite{HMM}.  Still they seem unlikely to show up
directly in the next generation of high energy collider experiments.
It is the indirect low energy effects induced by these CP-odd TGVs,
especially the electric-dipole-moment (EDM) of the neutron
\cite{MQ,Dju}, that will be more accessible to future experiments.  We
thus include in this paper estimates of the induced EDMs of the
neutron (and electron) in technicolor theories.

  The paper is organized as follows.  In section~2, we write down the
CP-odd TGV's, and the CP-violating dimension-four chiral lagrangian
operators, along with a table of the C, P and CP transformation
properties of all chiral lagrangian operators up to dimension four.
The connection between the CP-odd TGV's and the CP-violating chiral
lagrangian operators is made.  In section~3, we compute the
contribution to the CP-odd TGV form factors from a class of
CP-violating four-fermion operators induced by single ETC boson
exchange. The future high energy experimental detectability of these
TGVs is then discussed.  The zero momentum limit of the TGVs is
studied in connection with the CP-violating electroweak chiral
lagrangian operators and the $W$ boson EDM.  In section~4, the EDMs of
the electron and neutron, induced by the P- and CP-odd $WW\gamma$
effective interaction, are computed and compared to current
experimental limits.  Some other contributions to the EDMs of the
neutron and the electron are found to be less important.  We present
our conclusions in section~5.

\section{CP-Violating TGV's And The Chiral Lagrangian}

   We begin by writing down the CP-odd $WWV$ ($V=\gamma,Z$) vertices
in momentum space, following the notation of Ref.~\cite{Hag}:
\begin{eqnarray}
\Gamma^{\mu\nu\rho}_V (p_1,p_2,p) & = &
    i f_4^V(p^2) ( p^{\mu} g^{\rho\nu} + p^{\nu} g^{\rho\mu}) -
    f_6^V(p^2) \epsilon^{\rho\mu\nu\lambda} p_{\lambda} \nonumber \\ &
    & - \frac{f_7^V(p^2)}{\Lambda_{\chi}^2} (p_1 - p_2)^{\rho}
    \epsilon^{\mu\nu\lambda\sigma} p_{\lambda} (p_1 - p_2)_{\sigma},
\end{eqnarray}
where $p_1$, $p_2$ and $p = p_1 + p_2$ are the momenta carried by the
$W^{-}$, $W^{+}$ and $V$ bosons respectively.  The $W$ bosons are
taken to be on-shell. The form factors $f_4^V(p^2)$, $f_6^V(p^2)$ and
$f_7^V(p^2)$ could have imaginary parts if $p^2$ is above physical
thresholds. Since we will focus attention on effects from new physics
above $\Lambda_{\chi}$, we have defined $f_7^V(p^2)$ to be associated
with $\frac{1}{\Lambda_{\chi}^2}$, instead of $\frac{1}{m_W^2}$ as was
done in Ref.~\cite{Hag}.  We use the convention $\epsilon_{0123}=1$.

    The effective lagrangian for the CP-odd TGV's, which in momentum
space corresponds to the $p \rightarrow 0$ limit of Eq.~(2), reads
\cite{Hag}:
\begin{eqnarray}
 {{\cal L}_{WWV}^{\mbox{CP odd}} \over g_{WWV}} & = & -
 g_4^{V}W^{+}_{\mu}W^{-}_{\nu} ({\partial}^{\mu} V^{\nu} +
 {\partial}^{\nu} V^{\mu} ) + i {\tilde{\kappa}}_{V} W^{+}_{\mu}
 W^{-}_{\nu} {\tilde{V}}^{\mu\nu} \nonumber \\ & & +\, \frac{i
 {\tilde{\lambda}}_{V} }{\Lambda_{\chi}^2} W^{+}_{\mu\nu} W^{-\nu}_{\
 \ \ \rho} {\tilde{V}}^{\rho\mu},
\end{eqnarray}
where $\ W^{\pm}_{\mu\nu}={\partial}_{\mu}W^{\pm}_{\nu} -
{\partial}_{\nu}W^{\pm}_{\mu}$, $\ V_{\mu\nu}={\partial}_{\mu}V_{\nu}
- {\partial}_{\nu}V_{\mu}$, and ${\tilde{V}}_{\mu\nu}=\frac{1}{2}
{\epsilon}_{\mu\nu\rho\lambda} V^{\rho\lambda}$. The parameters
$g_{WW\gamma}$ and $g_{WWZ}$ are given by $g_{WW\gamma}= -e$ and
$g_{WWZ}= -e \cot \theta_W$, where $e$ is the electromagnetic coupling
constant, and $\theta_W$ is the Weinberg angle.  The TGV couplings
$g_4^V$, ${\tilde{\kappa}}_{V}$ and ${\tilde{\lambda}}_{V}$ are
related to the form factors by,
\begin{eqnarray}
f_4^V(0) & = & g_4^V \nonumber \\ f_6^V(0) & = & {\tilde{\kappa}}_{V}
- {\tilde{\lambda}}_{V} \frac{m_W^2} {\Lambda_{\chi}^2} \nonumber \\
f_7^V(0) & = & - \frac{1}{2} {\tilde{\lambda}}_{V}.
\end{eqnarray}
The electric dipole moment (EDM) $d_{W^+}$ and the magnetic quadrupole
moment $\tilde{Q}_{W^+}$ of the $W^+$ from physics above
$\Lambda_{\chi}$ are directly related to the TGV couplings by $d_{W^+}
=\frac{e}{2m_W}({\tilde{\kappa}_{\gamma}} + {\tilde{\lambda}_{\gamma}}
\frac{m_W^2}{\Lambda_{\chi}^2})$ and $\tilde{Q}_{W^+} =
-\frac{e}{m_W^2} ({\tilde{\kappa}_{\gamma}} -
{\tilde{\lambda}_{\gamma}} \frac{m_W^2}{\Lambda_{\chi}^2})$
respectively.

    The first term in Eq. (3) is P-even and CP-odd. The second and the
 third terms are both P- and CP-odd. The first two terms are of
 dimension four and correspond to certain dimension-four terms in the
 electroweak chiral lagrangian. The last term is of dimension six and
 is suppressed by a factor of ${\cal O}(\frac{p^2}{\Lambda_{\chi}^2})$
 relative to the CP-odd dimension-four terms. It will be neglected in
 this paper. The $g_4^{\gamma}$ term vanishes as a consequence of
 electromagnetic gauge invariance.

  We turn next to the electroweak chiral lagrangian.  This is an
effective low energy theory valid below the scale $\Lambda_{\chi}$.
It can be obtained from a more fundamental theory by integrating out
the (strongly interacting) heavy degrees of freedom above
$\Lambda_{\chi}$, and the high energy physics effects are then
contained in the coefficients of the chiral lagrangian operators.
Sixteen independent operators up to dimension four can be written
down~\cite{AW}, after using the equations of motion, for the gauged,
electroweak chiral $SU(2)_L \times SU(2)_R$ lagrangian.  The leading
operator can be obtained from the gauged scalar sector of the minimal
standard model by taking the physical Higgs boson mass to infinity.
It is given by,
\begin{eqnarray}
{\cal L}_0 & = & \frac{f^2}{4} \mbox{Tr} [(D_{\mu}U)^{\dag}
(D^{\mu}U)],
\end{eqnarray}
where $U \equiv \mbox{exp} (i \vec{\pi} \cdot \vec{\tau}/f)$ and
 $D_{\mu}U$ is its electroweak gauge covariant derivative.  The
 $\tau$'s are the three Pauli matrices.  Note that $U^{\dag}U=1$ and
 the dimensions of $U$ and $D_{\mu}U$ are zero and one respectively.
 The dimension-two operator ${\cal L}_0$ gives rise to the $W$ and $Z$
 masses with the relation $\frac{m_W^2}{m_Z^2}= \cos^2
 \theta_W$. Deviations from this relation can be described by the one
 other dimension-two operator, ${\cal L}_1^{'}$ of Ref.~\cite{ABL},
 directly related to the $T$ parameter.

The fourteen dimension-four chiral lagrangian operators are suppressed
by a factor of $\frac{1}{\Lambda_{\chi}^2}$ relative to the
dimension-two operators. The coefficients of these dimension-four
operators are then naturally of order $\frac{f^2}{\Lambda_{\chi}^2} =
\frac{N_f}{16 \pi^2}$ times factors arising from weak-isospin breaking
and CP violation(for the CP-violating operators).

   The three independent, CP-violating, dimension-four operators of
 the electroweak chiral lagrangian were designated ${\cal L}_{12}$,
 ${\cal L}_{13}$, and ${\cal L}_{14}$ in Ref.~\cite{AW}.  They are
 given by,
\begin{eqnarray}
{\cal L}_{12} & \equiv & 2 \, \alpha_{12} \, g\, Tr(T{\cal V}_{\mu})
                 Tr({\cal V}_{\nu}W^{\mu\nu}) \\ {\cal L}_{13} &
                 \equiv & \frac{1}{4} \, \alpha_{13} \, g\, g'\,
                 {\epsilon}^{\mu\nu\rho\sigma} \,
                 B_{\mu\nu}Tr(TW_{\rho\sigma}) \\ {\cal L}_{14} &
                 \equiv & \frac{1}{8} \, \alpha_{14} \, g^2 \,
                 {\epsilon}^{\mu\nu\rho\sigma} \, Tr(TW_{\mu\nu})
                 Tr(TW_{\rho\sigma}),
\end{eqnarray}
where $W_{\mu\nu}$, $B_{\mu\nu}$ are the gauge field strengths of
$SU(2)_L$ and $U(1)_Y$ respectively, and $g$ and $g'$ are their
corresponding gauge couplings. The operators $T$ and ${\cal V}_{\mu}$
live in the adjoint representation of $SU(2)_L$ and are singlets under
$U(1)_Y$. They are defined by $T \equiv U \tau_3 U^{\dag}$ and ${\cal
V}_{\mu} \equiv (D_{\mu} U) U^{\dag}$
\footnote{The operator field ${\cal V}_{\mu}$ used in this paper
 is the same as denoted by $V_{\mu}$ in Refs. \cite{ABL,AW}, it is
used here to avoid confusion with $V_{\mu}$ denoting the $\gamma$ or
$Z$ field.}.  The definitions adopted here for the coefficients
$\alpha_{12}$, $\alpha_{13}$ and $\alpha_{14}$ differ from those in
Ref.~\cite{AW} by the symmetry factors of $2$, $\frac{1}{4}$ and
$\frac{1}{8}$ respectively, in order to simplify the relations to the
TGV couplings below.

   It is interesting to observe that the operator ${\cal L}_{13}$ is
 the CP-violating analog of ${\cal L}_{1}$ of Refs.~\cite{ABL,AW},
 which is directly related to the weak-isospin symmetric $S$
 parameter. Similarly ${\cal L}_{14}$ is the CP-violating analog of
 ${\cal L}_{8}$, directly related to the weak-isospin-breaking $U$
 parameter.  The CP-conserving partner of ${\cal L}_{12}$ is the
 custodial-symmetry-breaking operator ${\cal L}_{11}$ of
 Ref.~\cite{AW}, which is the unique dimension-four P- and C-violating
 but CP-conserving operator ${\cal L}_{11} \equiv \alpha_{11} \, g
 \epsilon^{
\mu\nu\rho\lambda} \mbox{Tr} (T{\cal V}_{\mu}) \mbox{Tr}
    ({\cal V}_{\nu}W_{\rho\lambda})$.

   For convenience in relating the chiral lagrangian operators to the
electroweak gauge boson vertices, we give in Table~1 the C, P and CP
transformation properties of all chiral lagrangian operators up to
dimension four in the unitary gauge (setting the Goldstone-boson
fields to zero).

\vskip 0.3 in
\begin{center}
\begin{tabular}{|c|c|c|c|} \hline\hline
    & C & P & CP \\ \hline ${\cal L}_0$, ${\cal L}_1^{'}$, ${\cal
L}_1-{\cal L}_{10}$ & even & even & even \\ \hline ${\cal L}_{11}$ &
odd & odd & even \\ \hline ${\cal L}_{12}$ & odd & even & odd \\
\hline ${\cal L}_{13}$, ${\cal L}_{14}$ & even & odd & odd \\
\hline\hline
\end{tabular}
\end{center}
\vskip 0.2 in
{\small

   Table 1: The C, P and CP transformation properties of all the
sixteen $SU(2)_L \times U(1)_Y$ invariant chiral lagrangian operators
up to dimension four in unitary gauge.  The operators ${\cal L}_0$ and
${\cal L}_1^{'}$ have dimension two, and the operators ${\cal L}_1$
through ${\cal L}_{14}$ have dimension four. All these operators have
been listed in Ref.~\cite{AW}.

 }
\vskip .05 in
Note that the sixteen chiral lagrangian operators in general do not
have simple transformation properties separately under C and P, though
they can be classified as CP-even and CP-odd under the combined CP
operation.  In unitary gauge, however, they can all be classified as
C(P)-even or C(P)-odd, as shown in Table~1.  It can be seen from the
table that in unitary gauge, both ${\cal L}_{11}$ and ${\cal L}_{12}$
have unique transformation properties under C, P and CP.  Thus, these
two operators might be isolated and probed experimentally, if proper
observables are chosen and their size is large enough
\footnote{Experimental signatures of the CP-conserving operator ${\cal
L}_{11}$ have been studied in Ref.~\cite{Daw}. }.

 In unitary gauge, the CP-even chiral lagrangian operators reduce to
\mbox{two-,} three-, and four-gauge-boson vertices, whereas the
CP-odd ones generate only three- and four-gauge-boson vertices.  There
 are three independent CP-violating three-gauge-boson vertices,
 generated by ${\cal L}_{12}$, ${\cal L}_{13}$ and ${\cal L}_{14}$.
 They correspond to terms in the TGV effective lagrangian in Eq.~(3).
 The CP-odd TGV couplings are related to $\alpha_{12}$, $\alpha_{13}$
 and $\alpha_{14}$ by,
\begin{eqnarray}
{\tilde{\kappa}_Z} & = & \frac{e^2}{c^2} {\alpha}_{13} -
  \frac{e^2}{s^2} {\alpha}_{14} \\ {\tilde{\kappa}_{\gamma}} & = & -
  \frac{e^2}{s^2} ({\alpha}_{13} + {\alpha}_{14}) \\ g_4^Z & = & -
  \frac{e^2}{s^2c^2} {\alpha}_{12} \\ g_4^{\gamma} & = & 0 ,
\end{eqnarray}
where $s^2 = 1 - c^2 \equiv \sin^2 \theta_W$. The on-shell condition
for the $W$ bosons has been used in deriving the above relations.

  It is noted that there exists only one CP-odd four-gauge-boson
vertex at the dimension-four level.  It is generated by ${\cal
L}_{12}$.  (The CP-violating four-gauge-boson effective interaction
described by ${\cal L}_{14}$ is proportional to
$\epsilon_{\mu\nu\rho\sigma} W_1^{\mu} W_2^{\nu} W_1^{\rho}
W_2^{\sigma}$, thus vanishes.)  This C-odd and P-even (CP-odd)
effective $WWZ\gamma$ vertex is given by,
\begin{eqnarray}
{\cal L}_{WWZ\gamma} & = & i \alpha_{12} \frac{g^3e}{c} (W^{+ \mu}
 W^{-\nu} - W^{+ \nu} W^{-\mu} ) Z_{\mu} A_{\nu},
\end{eqnarray}
where $A_{\nu}$ denotes the photon field.  This vertex could be
interesting experimentally because of its uniqueness.  For example, it
could be studied at an $e\gamma$ collider via the reaction $e^- \gamma
\rightarrow \nu_e W^- Z$, which has been used as a probe for the
operator ${\cal L}_{11}$ in Ref.~\cite{Daw}.

\section{CP-Violating Four-Fermion Operators}

   We now proceed to estimate the above CP-violating TGV parameters in
technicolor theories, arising from the underlying ETC interactions.
We consider a one-family technicolor model, along with the three
quark-lepton generations.  The technifermions are taken to belong to
the fundamental representation of an $SU(N)_{TC}$ technicolor theory,
and transform like ordinary fermions under the standard model gauge
group.  We assume three different ETC scales, one for each family.
The ETC scale associated with the first family is much higher than
that associated with the second family, and the ETC scale for the
second family is much higher than that for the third.  For simplicity,
attention will first be restricted to the class of ETC models in which
the ETC gauge group commutes with the standard model gauge group.  We
will describe the CP violation in the ETC interactions in terms of a
set of CP-violating, $SU(N)_{TC} \times SU(3)_C \times SU(2)_L \times
U(1)_Y$ invariant four-fermion operators induced by single massive ETC
gauge boson exchange.

 It can be shown that all the single-ETC-boson-induced
four-technifermion operators which are $SU(N)_{TC} \times SU(3)_C
\times SU(2)_L \times U(1)_Y$ invariant are CP-conserving. It can also
be shown that all the CP-violating four-ordinary-fermion operators
involve fermions from at least one of the two light families, and are
thus suppressed by higher ETC scales.  It is only the CP-violating
four-fermion operators involving two technifermions and two ordinary
fermions from the heaviest family that are suppressed by the lowest
ETC scale and therefore have the largest CP-violating effects.  The
contribution of these four-fermion operators to the CP-odd TGV's turns
out to involve two mass insertions of the ordinary fermions. The
largest contribution will therefore come from the CP-violating
four-fermion operators consisting of two techniquarks and the ($t$,
$b$) doublet.  There are two of them, given by \cite{ABCH},
\begin{eqnarray}
{\cal L}_{4f}^1 & \equiv & ( \bar{q}_L \gamma^{\mu} Q_L ) [ \bar{Q}_R
      \gamma_{\mu} ( a_1 + a_1^{'} \, \tau_3 ) q_R ] + \mbox{h.c.} \\
      {\cal L}_{4f}^2 & \equiv & a_2 \, ( \bar{q}_R \gamma^{\mu} Q_R )
      ( \bar{Q}_R \gamma_{\mu} \tau_3 q_R ) + \mbox{h.c.}
\end{eqnarray}
where $q$ and $Q$ denote the ($t$, $b$) doublet and one doublet of
techniquarks respectively.  The $a$'s are in general complex, with
their magnitudes expected to be of order
$\frac{g_{ETC}^2}{m_{ETC}^2}$, where $g_{ETC}$ is the ETC coupling and
$m_{ETC}$ denotes the mass of the ETC boson.

  The operator ${\cal L}_{4f}^1$ is responsible for the generation of
quark masses in technicolor theories, as can be seen by a Fierz
rearrangement. When extended to include three families of quarks, it
gives rise to the CKM matrix with the Kobayashi-Maskawa (KM) phase
\cite{KM}.  Since CP symmetry is respected by TC interactions, the
techniquark condensates will naturally be real. It can then be shown
that ${\cal L}_{4f}^1$ with three quark families violates CP only via
the KM phase in the CKM matrix appearing in $W$ couplings to the
quarks, as well as in four-fermion operators involving both up-type
and down-type quarks. All CP-violating effects of ${\cal L}_{4f}^1$
(with three families) will then be suppressed by the small mixing
among the three quark-families and will not be considered here.

  We turn now to ${\cal L}_{4f}^2$. After the quarks and techniquarks
get their masses, we can rewrite the effective ETC lagrangian in terms
of the mass eigenstates of the fermion fields. It can then be seen
that ${\cal L}_{4f}^2$ violates CP conservation even with only one
family of quarks.  Though the CP phase in ${\cal L}_{4f}^2$ could be
rotated away by redefining the quark and techniquark fields, it will
reappear in other pieces of the effective lagrangian, such as the $W$
couplings to $t$, $b$ doublet and to techniquarks.  Physical
CP-violating effects are independent of this phase rotation.

  The operator ${\cal L}_{4f}^2$ contributes to the CP-odd TGV's via
the diagrams shown in Fig.~1. The techniquarks in Fig.~1 will of
course experience strong TC interactions.
  As is sometimes done in both QCD and technicolor theories,
one could model the strong interactions by employing dynamical mass
functions $\Sigma (p)$ for the single-loop of
  techniquarks and neglecting additional
TC interactions.  The $\Sigma (p)$'s can be determined by solving
Schwinger-Dyson gap equations, with their normalizations fixed by the
technipion decay constant $f$ (equivalently the $W$ and $Z$ masses)
\cite{PSHT}.

After a Fierz rearrangement, it can be seen that the CP-odd piece of
${\cal L}_{4f}^2$ consists of the product of a charge-changing
techniquark current and a charge-changing quark current.  The CP-odd
TGV's are induced by diagrams with the two $W$ bosons attached
separately to the quark and techniquark loops, and the $V$ ($\gamma$
or $Z$) boson attached to the quark loop, as in Fig.~1.  As ${\cal
L}_{4f}^2$ consists of two right-handed currents, two mass insertions
are needed on each loop.  The quark loop integral is then finite.
Since the techniquarks develop dynamical masses via the strong TC
interactions, with their magnitudes on the order of $\Lambda_{\chi}$
below the technifermion chiral symmetry breaking scale
$\Lambda_{\chi}$ and falling rapidly with momentum above
$\Lambda_{\chi}$, the techniquark loop integral will be highly
convergent at momentum scales just above $\Lambda_{\chi}$.

   The diagrams with the $V$ boson attached to the techniquark loop
are suppressed by a factor of ${\cal O} (\frac{m_t^2}{m_{TF}^2})$
relative to those with the $V$ boson attached to the quark loop as in
Fig.~1. This can be understood as follows: in the former case it is
the techniquark mass squared $m_{TF}^2$ that sets the infrared (IR)
scale, appearing in the denominator of the technifermion loop
integral; while in the latter it is the top quark mass squared $m_t^2$
that appears in the the denominator of the quark loop integral.  Thus
we need consider only the diagrams in Fig.~1.

   To simplify the computations, we use a momentum independent
 techniquark mass on the order of the technifermion chiral symmetry
 breaking scale $\Lambda_{\chi}$. This will give a good
 order-of-magnitude estimate of the techniquark loop integral, because
 the two techniquark mass insertions in the integral render this
 constant-mass computation at most logarithmic divergent, cut off at a
 scale $\Lambda_{CO}$ just above $\Lambda_{\chi}$.  As the logarithm
 is typically of order one, the constant mass estimate is expected to
 be the same order of magnitude as the dynamical mass calculation.

 For the incoming momentum $p$ of the $V$ ($\gamma$ or $Z$) boson
 below $\Lambda_{\chi}$, the evaluation of the two-loop diagrams in
 Fig.~1 for the CP-violating TGV's gives (taking the CKM matrix
 element $V_{tb}=1$),
\begin{eqnarray}
 f_6^Z(p^2) & = & \frac{N_C}{16 \pi^2} \frac{e^2}{s^2 c^2} \, \delta (
      ( -\frac{1}{2} + \frac{2}{3} s^2 ) I_1^{tbb}(p^2) + (
      \frac{1}{2}-\frac{4}{3} s^2) I_{1}^{btt}(p^2) ) \\
      f_6^{\gamma}(p^2) & = & \frac{N_C}{16 \pi^2} \frac{e^2}{s^2} \,
      \delta \, ( - \frac{2}{3} I_1^{tbb}(p^2) + \frac{4}{3}
      I_{1}^{btt}(p^2) ) \\ f_4^Z(p^2) & = & \frac{N_C}{16 \pi^2}
      \frac{e^2}{s^2 c^2} \, \delta \, ( \frac{1}{2} I_2^{tbb}(p^2) +
      \frac{1}{2} I_{2}^{btt}(p^2) ) \\ f_4^{\gamma}(p^2) & = & 0 ,
\end{eqnarray}
 where $N_C =3$ is the number of colors.  The $p^2$-dependent
 functions $I_{1,2}^{tbb}(p^2)$ and $I_{1,2}^{btt}(p^2)$ come from the
 quark loop integration and are given by,
\begin{eqnarray}
I_1^{tbb}(p^2) & = & \int_0^1 dx \int_0^{1-x} dy \,\, \frac{m_t m_b}
              {P(x,y)} \nonumber \\ I_2^{tbb}(p^2) & = & \int_0^1 dx
              \int_0^{1-x} dy \,\, (2x-1)\frac{m_t m_b } {P(x,y)},
\end{eqnarray}
where $P(x,y)$ is a polynomial in $x$ and $y$,
\begin{eqnarray}
P(x,y) & = & p^2 xy + m_W^2 (x+y)(1-x-y) - m_t^2(1-x-y) - m_b^2(x+y).
\end{eqnarray}
 The functions $I_{1}^{btt}(p^2)$ and $I_2^{btt}(p^2)$ can be obtained
 by exchanging $m_t$ and $m_b$ in the above expressions for
 $I_{1}^{tbb}(p^2)$ and $I_2^{tbb}(p^2)$ respectively.  The
 $p^2$-independent $\delta$ parameter arises from the techniquark loop
 integration and is a measure of CP violation at momentum scales just
 below $\Lambda_{\chi}$.

In the constant mass ($m_{TF} \sim {\cal O}(\Lambda_{\chi}$))
approximation, one is led to the simple expression for $\delta$,
\begin{eqnarray}
\delta & \simeq & \mbox{Im} a_2 \, N_{TC} \frac{m_{TF}^2}{4 \pi^2} \,
  \ln \frac{\Lambda_{CO}^2}{m_{TF}^2}.
\end{eqnarray}
Recall that $a_{2}$ is the coefficient of the CP-violating
four-fermion operator ${\cal L}_{4f}^{2}$ in Eq.(15), and that the
CP-odd piece of ${\cal L}_{4f}^{2}$ is proportional to the imaginary
part of $a_2$ (i.e. $\mbox{Im}a_2$). It is worth noting that this
techniquark loop computation can be viewed as generating the
CP-violating, dimension-four chiral lagrangian operator~\cite{ABCH},
\begin{equation}
  i {\delta}^{\prime} \bar{q}_R \tau_3 U^{\dag} (D_{\mu} U)
  \gamma^{\mu} q_R + \mbox{h.c.},
\end{equation}
where $q=(t,b)$ and $\delta = 2 \mbox{Im} \delta^{\prime}$.  The
 remaining part of the computation leading to Eqs.~(16--19) then takes
 place in the low energy chiral lagrangian framework.  Considering
 that the size of $\delta$ estimated later could be of order
 $10^{-2}$, this operator suggests the possibility of large
 CP-violating effects in the decay $t \rightarrow W^{+} + b$.

      For numerical evaluations, we take $p^2 = (200 \mbox{GeV})^2$,
the typical energy at which the LEP(200) collider operates, and
$m_b\simeq 5$ GeV.  Since $p^2$ is then well above $b\bar{b}$
threshold, the functions $I_{1,2}^{tbb}$ will have an imaginary part
as well as a real part.  Depending on the top quark mass and the
parameter $\delta$, the results of Table~2 are obtained.

\vskip 0.2 in
\begin{tabular}{crrrc} \hline\hline  \\
$m_t$ (GeV) & $f_6^Z \, (\times 10^{-4} \cdot \delta)$ & $f_6^{\gamma}
\, (\times 10^{-4} \cdot \delta)$ & $f_4^Z \, (\times 10^{-4} \cdot
\delta)$ & $f_4^{\gamma}$ \\ \hline \\ 130 & $-$ 1.2 + 2.6 i & $-$ 6.7
+ 3.8 i & 2.3 + 0.8 i & 0 \\ \\ 160 & $-$ 0.4 + 2.4 i & $-$ 3.9 + 3.5
i & 1.8 + 0.6 i & 0 \\ \\ 175 & $-$ 0.2 + 2.3 i & $-$ 3.1 + 3.4 i &
1.6 + 0.5 i & 0 \\ \\ 200 & 0.1 + 2.1 i & $-$ 2.2 + 3.2 i & 1.4 + 0.4
i & 0 \\ \hline\hline
\end{tabular}
\vskip .2 in
{\small

   Table 2: Size of the CP-violating TGV form factors from the
          four-fermion operator ${\cal L}_{4f}^2$ in a generic ETC
          model. The numbers in the table are computed at $p^2= (200
          \mbox{GeV})^2$ and should be multiplied by $10^{-4} \delta$.

 }
\vskip 0.2 in

  To determine the size of these CP-violating TGVs, we next estimate
$\delta$. The ETC scale associated with the top quark must be in the
range from several TeV to 10 TeV to be consistent with a top mass of
150 GeV or above \cite{ENKMRS,AT}.  For our estimate $m_{ETC}$ will be
taken to be 5 TeV.  We will assume that the ETC interactions are
relatively strong, with $g_{ETC}^2 \sim \frac{4 \pi^2}{N_{TC}}$
\footnote{This is roughly the critical strength for the four-fermion
interactions to trigger chiral symmetry breaking in an asymptotically
free gauge theory. See Ref.~\cite{FFI} for details.}.  In the absence
of a specific ETC model of CP violation, there is no obvious reason
for CP-violating interactions to be suppressed in the ETC sector.  In
this paper, we will assume maximal CP violation in extended
technicolor interactions. As a consequence, the CP-violating
four-fermion operator ${\cal L}_{4f}^2$ will have approximately equal
strengths in its CP-even piece and CP-odd piece, namely, $|\mbox{Im}
a_2| \simeq |\mbox{Re} a_2| \simeq \frac{1}{4}
\frac{g_{ETC}^2}{m_{ETC}^2}$
\footnote{The CP violation observed in the neutral Kaon system
will not be upset by maximal CP violation in ${\cal L}_{4f}^2$ because
of the small mixing between the third and the first two families.}.
As the constant techniquark mass is of order 1~TeV, the size of the
CP-violating parameter $\delta$ from new physics above
$\Lambda_{\chi}$ can be estimated as,
\begin{eqnarray}
\delta & \simeq & \mbox{Im} a_2 \, N_{TC} \frac{m_{TF}^2}{4 \pi^2} \,
           \ln   \frac{\Lambda_{CO}^2}{m_{TF}^2} \nonumber \\ &
     \simeq & \frac{g_{ETC}^2}{4m_{ETC}^2} \, N_{TC} \frac{m_{TF}^2}{4
     \pi^2} \, \ln \frac{\Lambda_{CO}^2}{m_{TF}^2} \nonumber \\ &
     \simeq & \frac{4 \pi^2/N_{TC}}{4m_{ETC}^2} N_{TC}
     \frac{m_{TF}^2}{4 \pi^2} \, \ln \frac{\Lambda_{CO}^2}{m_{TF}^2}
     \nonumber \\ & \simeq & \frac{1}{4} \frac{m_{TF}^2}{m_{ETC}^2}
     \cdot \, {\cal O}(1) \, \nonumber \\ & \simeq & 0.01
\end{eqnarray}
up to an overall minus sign. Recall that $\Lambda_{CO}$, $m_{TF}$
and $\Lambda_{\chi}$ are of the same order of magnitude.

    The CP-violating TGV form factors, at $p^2 = (200 \mbox{GeV})^2$
and with the top quark mass 175 GeV, can then be estimated (up to an
overall minus sign),
\begin{eqnarray}
f_6^Z ((200 \mbox{GeV})^2) & \simeq & (- 0.2 + 2.3 i) \times 10^{-6}
 \nonumber \\ f_6^{\gamma} ((200 \mbox{GeV})^2) & \simeq & (- 3.1 +
 3.4 i) \times 10^{-6} \nonumber \\ f_4^Z ((200 \mbox{GeV})^2) &
 \simeq & ( 1.6 + 0.5 i) \times 10^{-6} \nonumber \\ f_4^{\gamma} & =
 & 0
\end{eqnarray}
Note that the form factors are complex. They also vary with $p^2$,
though the order of magnitude remains unchanged for $p^2$ below
$\Lambda_{\chi}^2$.  The above size of the CP-odd TGV form factors in
technicolor theories is comparable to or larger than estimates made in
other extensions of the standard model \cite{HMM}.

 These form factors can be studied at $p\bar{p}$ colliders and the
 next generation of $e^+e^-$ colliders via, for example, measurements
 of CP-odd asymmetries of the two $W$-decays in the purely leptonic,
 purely hadronic and mixed channels \cite{Hag,CKP}.  The expected
 precision of the TGV form factors attainable at LEP(200) is no better
 than ${\cal O}(10^{-2})$.  It thus seems hopeless for the effects of
 these CP-violating TGVs to show up there in $e^+e^- \rightarrow
 W^+W^-$.  Even in other high energy processes including $e\gamma
 \rightarrow W\nu$ and $\gamma\gamma \rightarrow WW$ \cite{QBC}, where
 it has been suggested that a limit of ${\cal O}(10^{-3})$ on the form
 factors $\mbox{Im}f_6^V$ can be reached at energies of about 1~TeV,
 direct observation of the CP-violating TGVs seems unlikely.  The most
 stringent bounds currently available on the CP-odd TGVs come from
 measurements of low energy quantities, especially the EDM of the
 neutron, to which we turn in the next section.

  The $p \rightarrow 0$ limit of the TGVs is relevant for low energy
quantities like the EDM of the $W$ boson.  It also makes direct
contact with the chiral lagrangian coefficients $\alpha_{12}$,
$\alpha_{13}$ and $\alpha_{14}$.  From Eqs.~(9--12) and Eqs.~(16--19),
we have for on-shell $W$ bosons (up to an overall sign associated with
$\delta$)
\begin{eqnarray}
\alpha_{12} & = & \frac{N_C}{16 \pi^2} \, \delta \,
    (- \frac{1}{2} I_2^{tbb}(0) - \frac{1}{2} I_2^{btt}(0) ) \simeq -
    3.1 \times 10^{-6} \nonumber \\
\alpha_{13} & = & \frac{N_C}{16 \pi^2} \, \delta \,
   ( \frac{1}{6} I_1^{tbb}(0) - \frac{5}{6} I_1^{btt}(0)) \simeq - 1.6
    \times 10^{-6} \nonumber \\
\alpha_{14} & = & \frac{N_C}{16 \pi^2} \, \delta \,
  ( \frac{1}{2} I_1^{tbb}(0) - \frac{1}{2} I_1^{btt}(0) ) \simeq - 1.7
    \times 10^{-5}.
\end{eqnarray}
The numbers are for a top quark mass of $175 \mbox{GeV}$, with maximal
CP violation in the ETC interactions assumed.  Note that the
enhancement of $\alpha_{14}$ relative to $\alpha_{12}$ and
$\alpha_{13}$ arises from the presence of an infrared logarithm $\ln
\frac{m_b^2}{m_W^2}$ in the integral function $I_1^{tbb}(0)$ but not
in $I_1^{btt}(0)$ and $I_2^{tbb}(0) = I_2^{btt}(0)$, and the partial
cancelation between the two terms in the expression of $\alpha_{13}$
in Eq.~(26).  The appearance of the IR
logarithm can be seen by a simple power counting analysis, and it
occurs only in the limit $p \rightarrow 0$.
Of course, the four integral functions have the same
chirality suppression factor $\frac{m_t m_b}{m_W^2}$
so that they go to zero in the limit of vanishing $m_b$.
Apart from the factor
$\frac{1}{16 \pi^2} \sim \frac{f^2}{\Lambda_{\chi}^2}$, these
coefficients contain the expected CP-violating suppression factor
$\delta \simeq \mbox{Im} a_2 N_{TC} \frac{m_{TF}^2}{4\pi^2}
\ln \frac{\Lambda_{CO}^2}{m_{TF}^2} \simeq \frac{m_{TF}^2}
{4m_{ETC}^2}$
arising from integrating out heavy degrees of freedom above
$\Lambda_{\chi}$.  This $\delta$ factor is of course absent in
the CP-conserving chiral lagrangian operators.

   The TGV parameters $\tilde{\kappa}_{\gamma}$, $\tilde{\kappa}_Z$
and $g_4^Z$ are related to $\alpha_{12}$, $\alpha_{13}$ and
$\alpha_{14}$ by Eqs.~(9--12).  The P- and CP-odd TGV couplings
$\tilde{\kappa}_{\gamma}$ and $\tilde{\kappa}_Z$ directly measure the
electric-dipole-moment of the $W$ boson $d_{W^+} =
\frac{e}{2m_W}{\tilde{\kappa}_{\gamma}}$, and the analogous
weak-dipole-moment of the $W$ boson respectively. These TGV couplings
can readily be obtained from Eq.~(26) (up to an overall sign):
\begin{equation}
\tilde{\kappa}_Z  \simeq 6.9 \times 10^{-6},
\;\;\;\;\;\;\;\;
\tilde{\kappa}_{\gamma}  \simeq  7.8 \times 10^{-6},
\;\;\;\;\;\;\;\;
g_4^Z \simeq 1.7 \times 10^{-6},
\end{equation}
where the numbers are for a top quark mass of $175 \mbox{GeV}$ and
maximal CP violation in the ETC sector.  The $W$ boson EDM, $d_{W^+} =
\frac{e}{2m_W}{\tilde{\kappa}_{\gamma}}$, is then given by,
\begin{eqnarray}
d_{W^+} & = & \frac{3}{4\pi^2} \, e \, \sqrt{2}G_F \, m_W \, \delta \,
         (\, \frac{2}{3} I_1^{btt}(0) - \frac{1}{3} I_1^{tbb}(0) \, )
         \nonumber \\ & \simeq & 9.6 \times 10^{-20} \, \delta \;
         \mbox{$e$ cm} \; \simeq \; \pm \; 1 \times 10^{-21} \;
         \mbox{$e$ cm}.
\end{eqnarray}
This is much larger than the standard model prediction, a three-loop
effect less than $10^{-38} \; \mbox{$e$ cm}$, and is comparable to or
larger than its estimates in other models of CP violation \cite{EDMw}.

\section{The EDMs of the Neutron and the Electron}

    There exist various indirect low energy experimental bounds on the
CP-violating TGVs \cite{MQ,Dju}. The most stringent constraint of this
sort comes from the neutron EDM \cite{MQ}, which can be induced by the
P- and CP-odd $WW\gamma$ effective interaction. Describing the
strength of this $WW\gamma$ vertex by $\tilde{\kappa}_{\gamma}$, the
computation of the induced neutron EDM $d_n^{\gamma}$ contains an
ultraviolet logarithmic divergence. This means that the neutron EDM is
described by a separate (dimension-five) term in the effective chiral
lagrangian --- extended to include quark fields --- and that in the
absence of a high energy theory there is in principle no connection
between the two low energy parameters $\tilde{\kappa}_{\gamma}$ and
$d_n^{\gamma}$.

   One approach to the logarithmic divergence of the neutron EDM in
the absence of a high energy theory has been to regulate the high
energy physics responsible for the low energy P- and CP-odd $WW\gamma$
interaction by assigning a form factor to the $WW\gamma$
vertex \cite{MQ}.  The form factor was taken to be approximately
constant ($\sim \tilde{\kappa}_{\gamma}$) up to new physics scale
$\Lambda \gg m_W$, and falling rapidly for momentum above
$\Lambda$. Then the induced neutron EDM was found to be~\cite{MQ},
\begin{eqnarray}
\frac{d_n^{\gamma}}{e} & = &
         - \, \frac{1}{16\pi^2} \, g_A \, \xi \,  \sqrt{2}
               G_F m_N \, \tilde{\kappa}_{\gamma} \, ( \, \ln
               \frac{\Lambda^2}{m_W^2} \, + \, {\cal O} (1) \, ),
\end{eqnarray}
where $g_A=1.26$ is the axial-vector coupling constant measured in
neutron $\beta$-decay, $\xi \simeq 0.2$ is the average fraction of the
neutron momentum carried by a valence quark, $G_F$ is the Fermi
constant and $m_N$ is the mass of the nucleon.  Assuming $|\ln
\frac{\Lambda^2}{m_W^2} + {\cal O}(1)| > 1$ and requiring that the
induced neutron EDM be less than its direct experimental bound
$|d_n^{\gamma}| \leq 1.2 \times 10^{-25} \; \mbox{$e$ cm}$
\cite{EDMn}, it follows that
\begin{equation}
 |\tilde{\kappa}_{\gamma}| < 2 \times 10^{-4}.
\end{equation}
 Our low energy estimate, Eq.~(27), lies well within this bound,
indicating that maximal CP violation in the extended technicolor
interactions is allowed by current experiments.

  With a high energy theory such as technicolor in hand, a direct
connection between $\tilde{\kappa}_{\gamma}$ and $d_n^{\gamma}$ can in
principle be established.  By starting with the CP-violating
four-fermion operators arising from ETC interactions, an estimate can
be made of the $WW\gamma$ form factor and the neutron (and electron)
EDM.  As the CP-violating effects from ${\cal L}_{4f}^1$ (with three
families) are suppressed by the small mixing among the three quark
families, the dominant contribution to the $WW\gamma$ form factor and
the EDMs of the neutron and electron will come from the operator
${\cal L}_{4f}^2$ of Eq.~(15).

 A fermionic EDM can be induced from the P- and CP-odd $WW\gamma$
vertex by connecting the two $W$ boson lines to the fermion line.  The
$WW\gamma$ form factor is generated from ${\cal L}_{4f}^2$ via the
diagrams in Fig.~1 for arbitrary $W$ boson momentum.  As will be shown
shortly, the induced fermionic EDM is dominated by $W$ boson momenta
on the order of $m_t$ and below, and the diagrams with the quark and
the techniquark loops interchanged are suppressed by ${\cal
O}(\frac{m_t^2}{m_{TF}^2})$ relative to the diagrams shown in Fig.~1.
They will thus not be considered here.  The momentum flows of the
effective $WW\gamma$ vertex are shown in Fig.~2.

  We are now interested in the limit $p \rightarrow 0$, where $p$
denotes the incoming momentum of the photon. Keeping terms up to
quadratic in $p$, the momentum space structure of the $WW\gamma$
vertex is,
\begin{eqnarray}
\Gamma_{\lambda\mu\nu} (k,p) & \equiv &
   - \, \epsilon_{\lambda\mu\nu\alpha} \, p^{\alpha} \,
 \frac{1}{8\pi^2} \, \frac{e^2}{s^2} \, \delta \cdot ( \, I_0 (k^2) +
 k \cdot p \; I_1 (k^2) \, ),
\end{eqnarray}
where $k$ and $k+p$ are the incoming and outgoing momenta of the
 $W_{\nu}^{-}$ and $W_{\mu}^{-}$ bosons respectively, and $\delta$
 arises from integrating out the techniquarks and is given by the
 $k^2$ independent expression of Eq.~(22) for $|k^2| \ll m_{TF}^2$.
 The CKM matrix element $V_{tb}$ has been set to unity.  The integral
 functions $I_0(k^2)$ and $I_1(k^2)$ come from the quark loop integral
 and are defined as,
\begin{eqnarray}
I_0(k^2) & = & \int_0^1 dx \; \frac{m_t m_b (3x-2)}{-k^2 x (1-x) +
                 m_t^2 (1-x) + m_b^2 x} \nonumber \\ I_1(k^2) & = &
                 \int_0^1 dx \; \frac{m_t m_b (3x-2) x (1-x)}{[-k^2 x
                 (1-x) + m_t^2 (1-x) + m_b^2 x ]^2}.
\end{eqnarray}
In Eq.~(31), the form factor is defined for the $W$ bosons either on-
or off-shell and in the limit $p \rightarrow 0$.  Requiring the $W$
bosons to be on-shell, Eq.~(31) reduces to the $f_6^{\gamma}(p^2)$
term of Eq.~(2) in the limit of small photon momentum.

   It can be seen from Eq.~(32) that the $WW\gamma$ vertex falls
as $\frac{1}{-k^2}$ (up to a logarithm) for $-k^2$ above roughly
$m_t^2$.  Therefore the induced fermionic EDM converges rapidly when
the $W$ boson loop momentum increases above $m_t$, and the dominant
contribution to the fermion EDM is expected to come from the momentum
range $ - k^2 \ll m_{TF}^2$, which justifies the $k^2$-independent
expression for $\delta$ given by Eq.~(22).

  We now estimate the EDM of the electron.  Keeping terms
linear in the electron mass $m_e$, a unitary-gauge calculation of the
diagram in Fig.~3 gives,
\begin{eqnarray}
\frac{d_e^{\gamma}}{e} & = & - \, (\frac{1}{16\pi^2})^2 \, \frac{e^2}
   {s^2} \, 2 \sqrt{2} G_F \, m_e \, \delta \cdot
               h(\frac{m_t^2}{m_W^2}, \frac{m_b^2}{m_W^2}),
\end{eqnarray}
where $G_F$ is the Fermi constant, and the function $h(t,b)$ is given
by,
\begin{eqnarray}
h(t,b) & = & \sqrt{tb} \: \int_0^1 dx \: \frac{3x-2}{t(1-x) + bx -
        x(1-x)} \, \ln \frac{t(1-x) + bx}{x(1-x)}.
\end{eqnarray}
 For $m_t = 175 \mbox{GeV}$ and $m_b \simeq 5 \mbox{GeV}$,
$h(\frac{m_t^2}{m_W^2}, \frac{m_b^2}{m_W^2}) \simeq 0.24$. Assuming
maximal CP violation in the ETC sector (taking $|\delta| \simeq
0.01$ from Eq.~(24)), the EDM of the electron is estimated to
be (up to an overall minus sign associated with~$\delta$),
\begin{eqnarray}
d_e^{\gamma} & \simeq & - \, 1.2 \times 10^{-27} \, \delta \;
 \mbox{$e$ cm} \; \simeq \; - \, 1 \times 10^{-29} \; \mbox{$e$ cm}.
\end{eqnarray}
It is again worth noting that the computation of the electron EDM can
be viewed as a two-step process. The first step consists of
integrating out techniquarks above the technicolor chiral symmetry
breaking scale, thereby generating the operator of Eq.~(23).  The
second step is doing a two-loop calculation in the electroweak chiral
lagrangian framework.

 The technicolor result for $d_e^{\gamma}$ (Eq.~35) is to be compared
 with the standard model prediction of less than $10^{-37} \;
 \mbox{$e$ cm}$ \cite{SMe}, arising at the four-loop level and further
 suppressed by the small mixing effect among the three families of
 quarks.  In technicolor theories, the dominant contribution to the
 electron EDM involves only the third family of quarks and the heavy
 techniquarks, and it arises at two loops in the weak coupling
 expansion.  The current experimental limit on the electron EDM,
 $d_e^{\gamma} = (-2.7 \pm 8.3) \times 10^{-27} \; \mbox{$e$ cm}$
 \cite{EDMe}, is about three orders of magnitudes larger than the
 estimate in technicolor theories.  However, there is a good prospect
 for significant improvement over the present experimental limit using
 the $YbF$ molecule~\cite{PC}.

 We next turn to the neutron EDM.  Keeping terms linear in the light
quark masses, and neglecting the $W$ couplings connecting the first
and the third families of quarks, the EDM of the $u$ or $d$ quark can
be similarly computed from the diagram in Fig.~3. The result is,
\begin{eqnarray}
\frac{d_f^{\gamma}}{e} & = & I_3^f \, (\frac{1}{16\pi^2})^2 \,
                  \frac{e^2}{s^2} \, 4 \sqrt{2} G_F \, m_f \, \delta
               \cdot h(\frac{m_t^2}{m_W^2}, \frac{m_b^2}{m_W^2}),
\end{eqnarray}
where $f = u$ or $d \,$, and $I_3^u = \frac{1}{2}$ and $I_3^d = -
 \frac{1}{2}$ are the third components of the weak isospin of the $u$
 and $d$ quarks.

  As the neutron EDM is measured below the hadronic scale where the
chiral symmetry breaking effects of the light quarks become important,
the constituent quark masses of $m_u \simeq m_d \simeq \frac{m_N}{3}$
($m_N$ being the nucleon mass) are appropriate for use in Eq.~(36).
By use of the nonrelativistic quark model relation, $d_n^{\gamma} =
\frac{4}{3} d_d^{\gamma} - \frac{1}{3} d_u^{\gamma}$,
we then get \cite{neu},
\begin{eqnarray}
\frac{d_n^{\gamma}}{e} & = & - \, \frac{10}{9} \, (\frac{1}
        {16\pi^2})^2 \, \frac{e^2}{s^2} \, \sqrt{2} G_F \, m_N \,
        \delta \cdot
               h(\frac{m_t^2}{m_W^2}, \frac{m_b^2}{m_W^2}).
\end{eqnarray}
For numerical evaluation, we take $m_t = 175 \mbox{GeV}$ and $m_b
 \simeq 5 \mbox{GeV}$, and again assume maximal CP violation
in the ETC interactions ($|\delta| \simeq 0.01$).
 The neutron EDM is then estimated to be (up to an overall sign),
\begin{eqnarray}
d_n^{\gamma} & \simeq & - \, 1.2 \times 10^{-24} \, \delta \;
  \mbox{$e$ cm} \; \simeq \; - \, 1 \times 10^{-26} \; \mbox{$e$ cm}.
\end{eqnarray}
In the standard model, by contrast, the neutron EDM appears at the
three-loop-level, and is estimated to be less than $10^{-31} \;
\mbox{$e$ cm}$ \cite{SMn}, far below the current experimental bound of
$|d_n^{\gamma}| < 1.2 \times 10^{-25} \; \mbox{$e$ cm}$ \cite{EDMn}.
Since our estimate of the neutron EDM is near the current experimental
bound, further experimental improvement~\cite{Ra} on the limit of the
neutron EDM could help to reveal the precise nature of CP violation in
extended technicolor interactions.

   It is interesting to note that by using the value of
$\tilde{\kappa}_{\gamma}$ of Eq.~(27) to describe the strength of the
P- and CP-odd $WW\gamma$ vertex and computing the induced fermion EDM
with an ultraviolet cutoff $\Lambda$, as in the previous section, the
size of the resulting neutron and electron EDMs roughly agrees with
our dynamical calculations of Eqs.~(35) and (38) when simply setting
the logarithm $\ln \frac{\Lambda^2}{m_W^2}$ to 1.  This can be
understood easily if we recall that in the dynamical computation, the
cutoff $\Lambda$ should be replaced by roughly $m_t$, as can be seen
from the form factor of Eqs.~(31) and (32), and that $\ln
\frac{m_t^2}{m_W^2} \simeq 1.6$ for a top quark mass of 175 GeV.

   To conclude this section, we note that the CP-violating operator of
Eq.~(23) can be extended to include the three known families of quarks
and leptons, with the leptonic operators being generated by
integrating out the technileptons.  A fermion EDM can then be induced
by these CP-violating operators at one-loop in the weak-coupling
expansion (see Fig.~4). For the EDMs of the neutron and the electron,
however, the external fermions of Fig.~4 belong to the first family,
and the corresponding ETC scale is about $1000$ TeV. As this scale is
much higher than the ETC scale for the third family, the neutron and
electron EDMs induced by the diagrams of Fig.~4 will be smaller than
those induced by the P- and CP-odd $WW\gamma$ vertex of Fig.~3.

  The neutron EDM is induced by both types of diagrams in Fig.~4, with
the external fermion denoting a $u$ or $d$ quark. For the internal
fermions being the first family of quarks, and using the
nonrelativistic quark model relation $d_n^{\gamma} = \frac{4}{3}
d_d^{\gamma} - \frac{1}{3} d_u^{\gamma}$, we find,
\begin{eqnarray}
\left. \frac{d_n^{\gamma}}{e} \right|_{\mbox{Fig.~4}}
           & = & \frac{4}{9} \, \frac{\sqrt{2}G_F}{4\pi^2}
        \,   V_{ud} \, (5 m_u - m_d) \, \delta_1 ,
\end{eqnarray}
where $\delta_1$ characterizes the strength of the CP-violating
 operator Eq.~(23) for the first family of quarks.  Taking the ETC
 scale for the first family to be $1000$ TeV, and assuming maximal CP
 violation in the ETC interactions, $\delta_1$ can be estimated by
 integrating out the techniquarks. It is of order $10^{-7}$ or
 less. Using the constituent quark mass $m_u \simeq m_d \simeq
 \frac{m_N}{3}$, the induced neutron EDM is (up to an overall sign)
\begin{eqnarray}
\left. d_n^{\gamma} \right|_{\mbox{Fig.~4}}
 & \sim & 10^{-28} \; \mbox{--} \; 10^{-27} \; \mbox{$e$ cm}.
\end{eqnarray}
This is one to two orders of magnitudes smaller than the P- and CP-odd
$WW\gamma$--vertex--induced EDM of Eq.~(38).  It can be further shown
that if the internal quarks belong to the second and third families,
the induced neutron EDM from Fig.~4 is either in range $10^{-28} \;
\mbox{--} \; 10^{-27} \; \mbox{$e$ cm}$ or smaller, assuming all the
relevant CP-violating effective $W\bar{f_1}f_2$ vertices have the same
strengths. Therefore these one-loop contributions to the neutron EDM
can be safely neglected.

   The electron EDM is generated only by the first diagram in Fig.~4.
A straightforward computation gives,
\begin{eqnarray}
\left. \frac{d_e^{\gamma}}{e} \right|_{\mbox{Fig.~4}}
  & = & \frac{\sqrt{2}G_F}{4\pi^2} \, m_{\nu_e} \, \delta_e ,
\end{eqnarray}
where $m_{\nu_e}$ is the mass of the electron neutrino and $\delta_e$
characterizes the strength of the leptonic version of the CP-violating
operator of Eq.~(23) for the first family of leptons, generated by
integrating out the technielectron and technineutrino.  As the mass of
a Dirac-type electron neutrino is experimentally bounded to be at most
several eV, its contribution to the electron EDM can be neglected.

   A sizable value for $m_{\nu_e}$ is possible via the see-saw
mechanism if the right-handed neutrino gets a large Majorana mass.  In
this case, $m_{\nu_e}$ will be the off-diagonal Dirac mass term in the
Majorana neutrino mass matrix, and it could be assumed to have a value
on the order of the mass of its charged lepton partner.  For our
estimate, the ETC scale is taken to be $1000$ TeV, and the
technielectron and the technineutrino masses are taken to be $400$ GeV
and $200$ GeV respectively. Then $\delta_e \sim {\cal O}(10^{-8})$
with maximal CP violation in the ETC interactions.  Assuming
$m_{\nu_e} \sim 1$ MeV, the induced electron EDM is then
(up to an overall sign)
\begin{eqnarray}
\left. d_e^{\gamma} \right|_{\mbox{Fig.~4}}
 & \sim & 10^{-31} \; \mbox{$e$ cm}.
\end{eqnarray}
This is two orders of magnitudes smaller than the
$WW\gamma$-vertex-induced
value of Eq.~(35). Therefore we can neglect these one-loop
contributions to the electron EDM.

   So far we have restricted attention to the CP-violating
four-fermion operators induced by exchanging color- and electroweak-
neutral ETC bosons. CP-violating effects from an ETC sector which
has ETC bosons carrying nontrivial standard model quantum numbers
can and should be estimated. For example,
in a complete ETC theory, quarks and leptons as
 well as techniquarks and technileptons can be unified at higher
 energies via the exchange of Pati-Salam colored gauge
 bosons \cite{Hod,AT}, and new CP-violating electroweak invariant
 four-fermion operators could be generated. Most notable are the
 operators involving two techniquarks and two technileptons, where
 there are no mass-insertion suppressions when generating the CP-odd
 TGVs. These four-technifermion operators, however, are suppressed by
 the masses of the Pati-Salam bosons which could be above $1000$ TeV
 \cite{AT}, and their CP-violating effects can thus be neglected.

\section{Conclusion}

In this paper, the CP-odd $WW\gamma$ and $WWZ$ vertices have been
examined in technicolor theories. This has been done using effective
four-fermion interactions to describe the high energy ETC sector
responsible for the weak CP violation.  The technicolor interactions
respect the CP symmetry.  The dominant contribution to the CP-odd TGVs
comes from the operator ${\cal L}_{4f}^2$ (Eq.~(15)), which is induced
by color- and electroweak- neutral ETC boson exchange. This
CP-violating four-fermion operator consists of a techniquark doublet
and the ($t$,$b$) doublet.

  The size of the CP-odd $WW\gamma$ and $WWZ$ form factors in
technicolor theories is estimated to be of order $10^{-6}$ at center
of mass energy $200 \mbox{GeV}$ (Eq.~(25)), assuming that CP symmetry
is maximally violated in the ETC interactions.  This is much larger
than the minimal standard model prediction, and is comparable to or
larger than estimates made in other extensions of the standard model
\cite{HMM}.  Unfortunately the precision on the CP-violating TGV form
factors attainable at LEP(200) is at best of ${\cal O}(10^{-2})$, and
it is impossible for these CP-violating effects to show up in the next
generation of $e^+e^-$ experiments.  Even in high energy processes
like $e\gamma \rightarrow W\nu$ and $\gamma\gamma \rightarrow WW$ at
about 1~TeV, where the limit on the TGVs could reach ${\cal
O}(10^{-3})$~\cite{QBC}, direct observation of these CP-violating
vertices still seems unlikely.

  The coefficients of the three CP-violating, dimension-four
electroweak chiral lagrangian operators, defined in the $p \rightarrow
0$ limit, are estimated in technicolor theories to be of order
$10^{-6} \, \mbox{--} \, 10^{-5}$ (Eq.~(26)), about three orders of
magnitudes smaller than the CP-conserving, dimension-four (custodial
symmetry preserving) ones estimated in Ref.~\cite{AW}.
This can be understood as follows. The
dimension-four CP-conserving (custodial symmetry preserving) chiral
lagrangian operators are generated by integrating out the strongly
interacting technifermions, and are independent of the higher scale
ETC interactions. Their size is therefore of order
$\frac{f^2}{\Lambda_{\chi}^2} \simeq \frac{N_f}{16\pi^2}$.  The
CP-violating chiral lagrangian operators, however, originate from the
high energy (CP-nonconserving) ETC interactions, and are naturally
suppressed by $\frac{\Lambda_{\chi}^2}{m_{ETC}^2}$ relative to the
CP-conserving operators. In the types of ETC models considered in this
paper, there is in addition the chirality suppression factor
$\frac{m_t m_b}{m_W^2}$ in the three CP-violating chiral lagrangian
operators.  The connection between the CP-odd TGV couplings and
CP-violating chiral lagrangian operators is made in Eqs.~(9-12).

  Severe constraints on the CP-odd TGVs can be obtained from low
energy quantities like the electric- and weak-dipole moments of the
quarks and leptons \cite{MQ,Dju}.  Among them the neutron EDM supplies
the most stringent bound. Consistency between the current experimental
limit on the neutron EDM and its induced value from a P- and CP-odd
$WW\gamma$ vertex requires $|\tilde{\kappa}_{\gamma}| < 2 \times
10^{-4}$.  In technicolor theories, the chiral lagrangian parameter
$\tilde{\kappa}_{\gamma}$ is estimated to be
$|\tilde{\kappa}_{\gamma}| \sim {\cal O}(10^{-5})$. This corresponds
to an EDM of the $W$ boson of ${\cal O}(10^{-21}) \; \mbox{$e$ cm}$,
and lies comfortably within the above indirect experimental bound.
The $W$ EDM in technicolor theories is much larger than the standard
model expectation $< 10^{-38} \, \mbox{$e$ cm}$ \cite{EDMw}, and is
comparable to or larger than its size in other models of CP violation.

The P- and CP-odd $WW\gamma$ vertex for off-shell $W$ bosons can be
computed in technicolor theories, and used to estimate the EDMs of
the neutron and the electron. The $WW\gamma$ vertex falls like
$\frac{1}{k^2}$ for the $W$ boson momentum $k$ above $m_t$, and a
finite expression can be written down for the induced fermion EDM. The
EDM of the electron is estimated to be as large as ${\cal O}(10^{-29})
\; \mbox{$e$ cm}$, three orders of magnitudes below the present
experimental limit.  Significant improvement over the current
experimental limit on the electron EDM is expected in the near future
using new experimental methods \cite{PC}.  The neutron EDM in
technicolor theories can be as large as ${\cal O}(10^{-26}) \;
\mbox{$e$ cm}$, not far from the current experimental bound of less
than $1.2 \times 10^{-25} \, \mbox{$e$ cm}$.  Future experimental
improvements \cite{Ra} on the limit of the neutron EDM could help to
reveal the precise nature of CP violation in extended technicolor
interactions.

  As the ETC scale for the first fermion family is two orders of
magnitudes higher than for the third family, the induced EDMs of the
neutron and electron from one-loop diagrams of Fig.~4 turn out to be
one to two orders of magnitudes smaller than induced from the P- and
CP-odd $WW\gamma$ vertex (Fig.~3) --- a two-loop effect in the weak
coupling expansion.  Because of the relatively low ETC scale
associated with the third family of quarks, it is also suggested that
potentially large CP-violating effects could be observed in the decay
$t \rightarrow b + W^+$.

\section*{Acknowledgments}

 We like to thank W.~Marciano and S.~Selipsky for helpful
conversations, and J.~Liu for pointing out to us Ref.~\cite{EDMw}.
This work is supported by a DOE grant DE-AC02ERU3075.


\begin{thebibliography}{10}
\bibitem{neu}
A different method, making use of the relation between the nucleon
matrix elements of the $u$ and $d$ quark EDM amplitude and the
$\beta$-decay axial current matrix elements, was employed in
Ref.~\cite{MQ} to obtain the neutron EDM from the perturbative
expression of the quark EDM.  This method was used to obtain the
estimate of Eq.~(29). It would give a neutron EDM
about half of the size estimated in Eq.~(38).
\end{thebibliography}
\end{document}